\documentclass[%
 aip,
 amsmath,amssymb,
 reprint,%
]{revtex4-1}

\usepackage{graphicx}
\usepackage{dcolumn}
\usepackage{bm}

\usepackage[utf8]{inputenc}
\usepackage[T1]{fontenc}
\usepackage{mathptmx}
\usepackage{etoolbox}
\usepackage{xcolor}

\makeatletter
\def\@email#1#2{%
 \endgroup
 \patchcmd{\titleblock@produce}
  {\frontmatter@RRAPformat}
  {\frontmatter@RRAPformat{\produce@RRAP{*#1\href{mailto:#2}{#2}}}\frontmatter@RRAPformat}
  {}{}
}%
\makeatother
\begin{document}


\title{Optimal Binning for Small-Angle Neutron Scattering Data Using the Freedman-Diaconis Rule}
\author{Jessie E. An}
\author{Chi-Huan Tung}
\author{Changwoo Do}
\author{Wei-Ren Chen}
\affiliation{Neutron Scattering Division, Oak Ridge National Laboratory, Oak Ridge, 37831, Tennessee, United States}

\date{\today}

\begin{abstract}
Small-Angle Neutron Scattering (SANS) data analysis often relies on fixed-width binning schemes that overlook variations in signal strength and structural complexity. We introduce a statistically grounded approach based on the Freedman–Diaconis (FD) rule, which minimizes the mean integrated squared error between the histogram estimate and the true intensity distribution. By deriving the competing scaling relations for counting noise ($\propto h^{-1}$) and binning distortion ($\propto h^{2}$), we establish an optimal bin width that balances statistical precision and structural resolution. Application to synthetic data from the Debye scattering function of a Gaussian polymer chain demonstrates that the FD criterion quantitatively determines the most efficient binning, faithfully reproducing the curvature of $I(Q)$ while minimizing random error. The optimal width follows the expected scaling $h_{\mathrm{opt}} \propto N_{\mathrm{total}}^{-1/3}$, delineating the transition between noise- and resolution-limited regimes. This framework provides a unified, physics-informed basis for adaptive, statistically efficient binning in neutron scattering experiments.
\end{abstract}

\maketitle


\section{Introduction}
\label{sec:1}

Small-Angle Neutron Scattering (SANS) is one of the most powerful experimental techniques for probing nanoscale structures in soft matter, polymers, and complex fluids \cite{Lindner2002}. By measuring the scattered neutron intensity as a function of the momentum transfer $Q$, SANS provides direct information on characteristic length scales from a few to hundreds of nanometers. Despite its versatility, SANS measurements often suffer from intrinsically low scattering intensities, particularly at high $Q$ or for dilute samples. Because each detector pixel collects only a small number of neutrons, the resulting data are statistically sparse, and the reduction of counting noise through proper binning becomes a crucial step in data analysis.

In current SANS practice, however, the binning of detector data is rarely treated as a quantitatively optimized process. A single, fixed binning scheme is often applied to all samples regardless of their scattering contrast or structural complexity. Yet, scattering profiles $I(Q)$ from different materials can vary greatly in shape and smoothness due to differences in composition, form factor, and interaction potential. Using the same bin width for all systems fails to account for these variations: some datasets may become over-binned, losing fine structural features, while others may be under-binned, retaining unnecessary statistical noise. At present, there is no established binning strategy that simultaneously preserves essential structural information and optimizes measurement efficiency. Developing such a criterion—one that balances statistical precision and expressive resolution—is therefore an important step toward consistent, high-fidelity SANS data reduction.

A statistically grounded approach to binning was established by Freedman and Diaconis \cite{Freedman1981}, who derived the optimal bin width for histogram-based density estimation by minimizing the mean integrated squared error between the histogram and the true underlying probability density. Their analysis yields the well-known Freedman-Diaconis (FD) rule,
\begin{equation}
h = 2\, \mathrm{IQR}\, n^{-1/3},
\end{equation}
where $\mathrm{IQR}$ is the interquartile range and $n$ is the number of data points. The FD rule provides a mathematically rigorous balance between resolution and statistical fluctuation: narrow bins increase variance by amplifying noise, whereas wider bins reduce variance but introduce bias by oversmoothing. Its simple analytic form, robustness against outliers, and independence from any assumed model have made the FD rule a universal reference for data-based histogram construction. It has been successfully applied in a broad range of disciplines, including imaging \cite{Gonzalez2008}, genomics \cite{Li2011}, astronomy \cite{Ivezic2014}, and signal processing \cite{Knuth2019}, but, to date, it has not been systematically implemented for SANS data analysis.

In contrast to heuristic rules, the FD method rests on first principles of statistical estimation. It defines a clear scaling relation between sample size and histogram resolution, establishing a deterministic link between data statistics and binning precision. Because of these advantages, it provides a natural foundation for developing a standardized and efficient SANS binning procedure.

In this project, we develop a binning strategy for SANS data based on the Freedman-Diaconis rule. The goal is to test and demonstrate how this statistically grounded approach can be applied to real SANS datasets to achieve a balance between resolution and noise, enabling consistent and reproducible data reduction across different instruments and materials. The focus of this work is on the practical implementation and validation of the FD method for neutron scattering data, without considering higher-order effects such as inter-pixel correlations. This study represents an initial step toward establishing a quantitative, physics-informed framework for optimal binning in SANS experiments.

\section{Methods}
\label{sec:2}

In scattering experiments, the central observable is the intensity distribution $I(Q)$ measured over the momentum-transfer range $[Q_{\min}, Q_{\max}]$, with $L = Q_{\max} - Q_{\min}$ defining the coherent window. A standard measure of the deviation between the histogram representation $I_h(Q)$ with bin width $h$ and the ground truth $I(Q)$ is given by the normalized $L_2$ deviation~\cite{strang2006linear}. Freedman and Diaconis assessed the quality of a binning scheme by minimizing the squared deviation between the histogram estimate and the underlying distribution in the sense of the $L_2$ norm~\cite{Freedman1981}. For a histogram estimate $I_{h}(Q)$ obtained with bin width $h$ and a ground-truth distribution $I(Q)$, the $L_{2}$ norm is
\begin{equation}
\| I_{h} - I \|_{2}^{2} 
= \int_{Q_{\min}}^{Q_{\max}} \bigl(I_{h}(Q) - I(Q)\bigr)^{2} \, dQ ,
\end{equation}
where $[Q_{\min},Q_{\max}]$ is the experimentally accessible scattering range of length $L = Q_{\max} - Q_{\min}$. Optimizing $h$ therefore amounts to balancing statistical fluctuations that scale as $h^{-1}$ against binning distortion that scales as $h^{2}$. In the following, we derive these two contributions explicitly.

We consider a detector that divides the $Q$-range $[Q_{\min}, Q_{\max}]$ into $K$ equal-width bins, each of size
\begin{equation}
h = \frac{Q_{\max} - Q_{\min}}{K}.
\end{equation}
Let $N_k$ denote the number of detected counts in bin $k$, with the total count given by
\begin{equation}
N_{\text{total}} = \sum_{k=1}^{K} N_k.
\end{equation}
Assuming that the measured intensity $I(Q_k)$ is proportional to $N_k$, we define the reconstructed intensity in bin $k$ as
\begin{equation}
I_k = C_I \, \frac{Q_{\max} - Q_{\min}}{h} \, \frac{N_k}{N_{\text{total}}},
\label{eq:4}
\end{equation}
where $C_I$ is a proportionality constant independent of binning and counting.  

Because neutron detection obeys Poisson statistics~\cite{Knoll2010}, the counting uncertainty in each bin scales as $(\Delta I_k^{\text{count}})^2 \propto N_k$, leading to
\begin{equation}
(\Delta I_k^{\text{count}})^2 = C_I^2 \, \frac{(Q_{\max} - Q_{\min})^2}{h^2} \, \frac{N_k}{N_{\text{total}}^2}.
\end{equation}
We approximate the expected number of counts in bin $k$ by
\begin{equation}
N_k \approx N_{\text{total}} \, \frac{\bar{I}_k \, h}{\int_{Q_{\min}}^{Q_{\max}} I(Q) \, dQ},
\end{equation}
where $\bar{I}_k$ is the bin-averaged intensity,
\begin{equation}
\bar{I}_k = \frac{1}{h} \int_{Q_k - h/2}^{Q_k + h/2} I(Q) \, dQ.
\end{equation}
Substituting this expression into the counting variance yields
\begin{equation}
(\Delta I_k^{\text{count}})^2 \sim C_I^2 \, \frac{(Q_{\max} - Q_{\min})^2}{h} \, \frac{\bar{I}_k}{N_{\text{total}} \int_{Q_{\min}}^{Q_{\max}} I(Q) \, dQ}.
\end{equation}
Averaging over all bins, and using the approximation
\begin{equation}
\frac{1}{K} \sum_{k=1}^{K} \bar{I}_k \sim \frac{1}{Q_{\max} - Q_{\min}} \int_{Q_{\min}}^{Q_{\max}} I(Q) \, dQ,
\end{equation}
we obtain the average squared counting error:
\begin{equation}
\left\langle (\Delta I^{\text{count}})^2 \right\rangle = \frac{C_I^2 \, (Q_{\max} - Q_{\min})}{N_{\text{total}} \, h}.
\label{eq:DeltaI_count}
\end{equation}
This result shows that the counting noise contribution scales inversely with bin width $h^{-1}$, reflecting that finer binning increases statistical fluctuations.

In addition to counting uncertainty, finite binning introduces a systematic distortion because the continuous intensity $I(Q)$ within each interval is represented by a single number. Let $Q_k$ be the center of the $k$th bin of width $h$. Define the bin average and the pointwise deviation within the bin as
\begin{equation}
\bar I_k = \frac{1}{h}\int_{Q_k-\frac{h}{2}}^{Q_k+\frac{h}{2}} I(Q)\,dQ,
\end{equation}
and
\begin{equation}
\Delta I_k^{\mathrm{bin}}(Q) = I(Q) - \bar I_k.
\label{eq:def_DeltaIbin(Q)}
\end{equation}
The within-bin mean squared binning error is then
\begin{equation}
\bigl(\Delta I_k^{\mathrm{bin}}\bigr)^2
= \frac{1}{h}\int_{Q_k-\frac{h}{2}}^{Q_k+\frac{h}{2}}
\bigl(\Delta I_k^{\mathrm{bin}}(Q)\bigr)^2\,dQ.
\label{eq:def_bin_mse}
\end{equation}

To evaluate this error, we expand $I(Q)$ about $Q_k$ using $x = Q - Q_k$:
\begin{equation}
I(Q_k + x) = I_k + I'_k x + \frac{1}{2} I''_k x^2 + \frac{1}{6} I'''_k x^3 + \frac{1}{24} I^{(4)}_k x^4 + \mathcal{O}(x^5).
\label{eq:expansion_I(Q_k)}
\end{equation}
The bin average can then be expressed as
\begin{equation}
\bar I_k = I_k + \frac{h^2}{24} I''_k + \frac{h^4}{1920} I^{(4)}_k + \mathcal{O}(h^6).
\label{eq:expansion_Ibar}
\end{equation}
Substituting \eqref{eq:expansion_I(Q_k)} and \eqref{eq:expansion_Ibar} into \eqref{eq:def_DeltaIbin(Q)} and computing the mean squared deviation \eqref{eq:def_bin_mse}, we obtain
\begin{equation}
\bigl(\Delta I_k^{\mathrm{bin}}\bigr)^2
= \frac{h^2}{12}\,(I'_k)^2
+ h^4\!\left(\frac{1}{240}\,I'_k I'''_k+\frac{1}{720}\,(I''_k)^2\right)
+\mathcal{O}(h^6).
\end{equation}
The leading scaling behavior is therefore $\bigl(\Delta I_k^{\mathrm{bin}}\bigr)^2 = \mathcal{O}(h^2)$ with leading coefficient $(I'(Q_k))^2/12$.  
Averaging over $K$ equal-width bins covering $[Q_{\min}, Q_{\max}]$, we find
\begin{equation}
\begin{split}
\left\langle (\Delta I^{\mathrm{bin}})^2 \right\rangle
&= \frac{1}{K}\sum_{k=1}^{K}\left(\Delta I_k^{\mathrm{bin}}\right)^2 \\
&= \frac{h^2}{12\,L}\int_{Q_{\min}}^{Q_{\max}} \left[I'(Q)\right]^2\,\mathrm{d}Q
\;+\;\mathcal{O}(h^4),
\end{split}
\label{eq:DeltaI_bin}
\end{equation}

Eq.~\eqref{eq:DeltaI_bin} shows that the binning distortion scales quadratically with bin width $h^2$, as narrower bins more accurately capture the curvature of $I(Q)$.

Combining the two contributions from counting noise [Eq.~\eqref{eq:DeltaI_count}] and binning distortion [Eq.~\eqref{eq:DeltaI_bin}], we obtain the total mean squared deviation:
\begin{equation}
\begin{split}
\left\langle (\Delta I)^2 \right\rangle 
&= \frac{1}{K} \sum_{k=1}^{K} 
\left[ 
\left( \Delta I_k^{\text{count}} \right)^2 
+ 
\left( \Delta I_k^{\text{bin}} \right)^2 
\right] \\
&= 
\left\langle (\Delta I^{\text{count}})^2 \right\rangle 
+ 
\left\langle (\Delta I^{\text{bin}})^2 \right\rangle.
\end{split}
\label{eq:DeltaI_total}
\end{equation}

Minimizing Eq.~\eqref{eq:DeltaI_total} with respect to $h$ yields the optimal bin width, providing the Freedman--Diaconis scaling relationship between noise and resolution for statistically efficient SANS data binning.

Eq.~\eqref{eq:DeltaI_total} shows that the total deviation arises from two competing contributions: a counting term that scales inversely with the bin width, $h^{-1}$, and a binning distortion term that scales quadratically with the bin width, $h^{2}$. The first term reflects the stochastic nature of neutron detection and is governed by the average intensity $\bar I_k$, which characterizes the local coherent scattering strength of the system. The second term depends on the local derivative $I'(Q)$, which quantifies the rate of variation of the scattering intensity with momentum transfer and therefore captures the local structural expressiveness of the material. Together, $\bar I_k$ and $I'(Q)$ describe the essential characteristics of the measured intensity profile, namely the overall scattering power and its spatial variability. The interplay between these two terms establishes a natural balance between statistical precision and structural resolution. Minimizing Eq.~\eqref{eq:DeltaI_total} thus defines an optimal bin width that adapts automatically to the intrinsic smoothness and signal strength of each $I(Q)$, providing a unified and physically grounded criterion for determining the most effective binning strategy across different SANS samples and measurement conditions.
 
\section{Results and Discussion}
\label{sec:3}

Following the formulation in Section~\ref{sec:2}, we apply the Freedman--Diaconis binning framework to a synthetic SANS dataset to demonstrate how the total deviation in Eq.~\eqref{eq:DeltaI_total} governs the balance between statistical noise and structural fidelity. The synthetic scattering intensity was generated using the Debye scattering function of a Gaussian polymer chain~\cite{Debye1947}, which provides a continuous and analytically defined reference profile $I(Q)$ over the momentum-transfer range $[Q_{\min}, Q_{\max}]$. This model is ideally suited for evaluating binning strategies because the intensity decreases monotonically with $Q$, exhibiting a well-characterized curvature and smooth derivative that make it sensitive to both under- and over-binning effects.

\begin{figure}[h]
    \centering
    \includegraphics[width=1\linewidth]{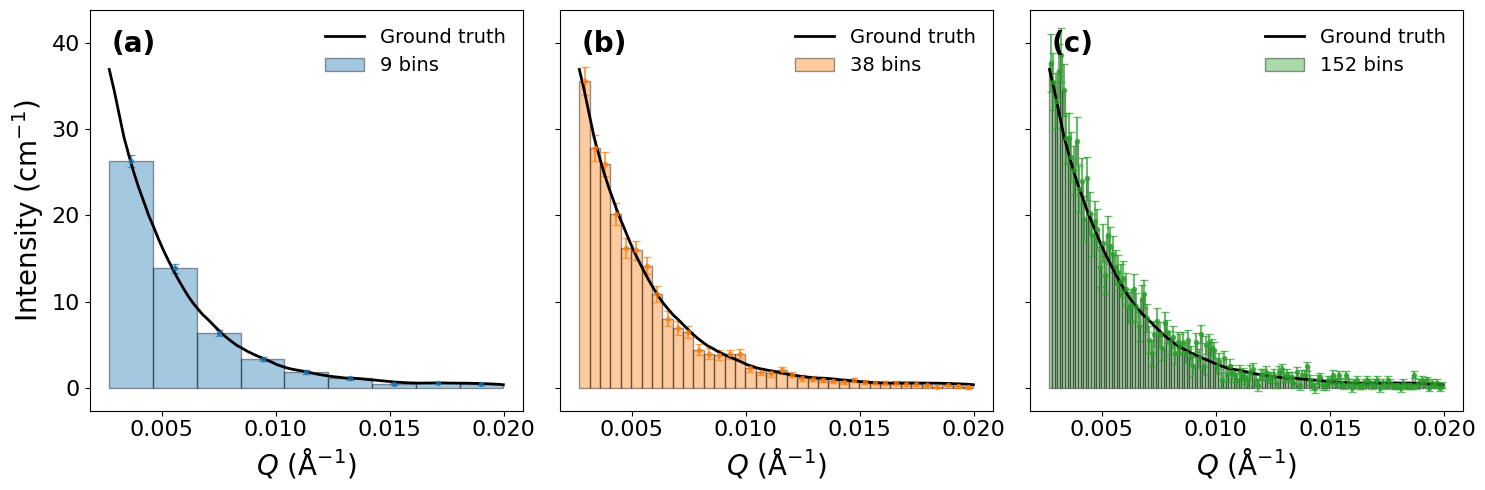}
     \caption{Reconstruction of a synthetic SANS intensity profile generated from the Debye scattering function of a Gaussian polymer chain~\cite{Debye1947}. The analytical $I(Q)$ (gray line) serves as the reference against which histograms obtained using different bin numbers are compared. (a) Coarse binning ($K = 10$) yields a smooth intensity profile with low statistical noise but significant distortion due to averaging over wide intervals. (b) Optimal binning ($K_{\mathrm{opt}} = 38$) minimizes the total mean-squared deviation in Eq.~\eqref{eq:DeltaI_total}, accurately reproducing the curvature and overall decay of the Debye function. (c) Over-binning ($K = 120$) produces strong statistical fluctuations dominated by counting noise. The three panels collectively demonstrate how the balance between counting variance ($\propto h^{-1}$) and binning distortion ($\propto h^{2}$) determines the optimal bin width, which preserves the intrinsic structural features of the scattering profile while minimizing random error.}
    \label{fig:1}
\end{figure}

Fig.~\ref{fig:1} presents the reconstructed intensity profiles obtained from simulated neutron counts using three representative bin widths. The total count statistics and scattering range were held fixed, while the number of bins $K$ was systematically varied. For each binning configuration, the intensity histogram $I_{h}(Q)$ was calculated from Eq.~\eqref{eq:4}, and the total mean-squared deviation $\langle (\Delta I)^2 \rangle$ was evaluated from Eq.~\eqref{eq:DeltaI_total}. The competing scaling of the two error components—$h^{-1}$ for counting noise and $h^{2}$ for binning distortion—yields a distinct minimum in $\langle (\Delta I)^2 \rangle$ as a function of bin width $h$. From these calculations, the optimal number of bins was found to be $K_{\mathrm{opt}} = 38$, corresponding to an intermediate resolution that minimizes the combined contribution of statistical and systematic errors.

Fig.~\ref{fig:1}(a) shows the reconstructed intensity using a coarse binning configuration ($K = 10$). The large bin width suppresses statistical fluctuations, resulting in a smooth curve, but the averaging over wide intervals obscures local variations in $I(Q)$. The curvature and slope of the Debye profile are poorly reproduced, indicating that the binning distortion term in Eq.~\eqref{eq:DeltaI_bin} dominates the total error. 

Fig.~\ref{fig:1}(b) displays the reconstruction obtained using the optimal bin width ($K_{\mathrm{opt}} = 38$). In this configuration, the histogram accurately follows the analytical Debye function across the entire $Q$ range, capturing both the gradual decay and the subtle curvature at intermediate $Q$. The variance and bias contributions are balanced such that $\langle (\Delta I)^2 \rangle$ reaches its minimum. This configuration defines the most efficient statistical representation of the data, preserving essential structural information without introducing spurious fluctuations. The agreement between the histogram and the analytical reference confirms that the FD criterion provides a quantitative and transferable measure of binning quality.

Fig.~\ref{fig:1}(c) illustrates the case of over-binning ($K = 120$). Here, each bin contains only a small number of neutron counts, and the resulting intensity exhibits pronounced pointwise fluctuations caused by Poisson noise. Although fine binning increases apparent resolution, it amplifies statistical variance and introduces artificial oscillations that do not correspond to any physical feature of the model. In this regime, the $h^{-1}$ term in Eq.~\eqref{eq:DeltaI_count} dominates, leading to a rapid increase in total deviation. 

The behavior observed across the three panels reflects the fundamental competition between statistical precision and resolution inherent to histogram-based data analysis. The counting term depends primarily on the average intensity $\bar{I}_k$, representing the local coherent scattering strength, while the binning distortion term depends on the local slope $I'(Q)$, reflecting the structural expressiveness of the signal. Because each material system exhibits its own characteristic $I(Q)$ profile and derivative, the resulting optimal binning number $K_{\mathrm{opt}}$ naturally adapts to the intrinsic smoothness and signal strength of the sample. In the present synthetic case, $K_{\mathrm{opt}} = 38$ represents the point of minimal total deviation, defining the optimal trade-off between noise suppression and feature preservation. This result provides a quantitative example of how the Freedman--Diaconis framework can be used to design a statistically consistent and physically meaningful binning strategy for SANS data.

\begin{figure}[h]
    \centering
    \includegraphics[width=1\linewidth]{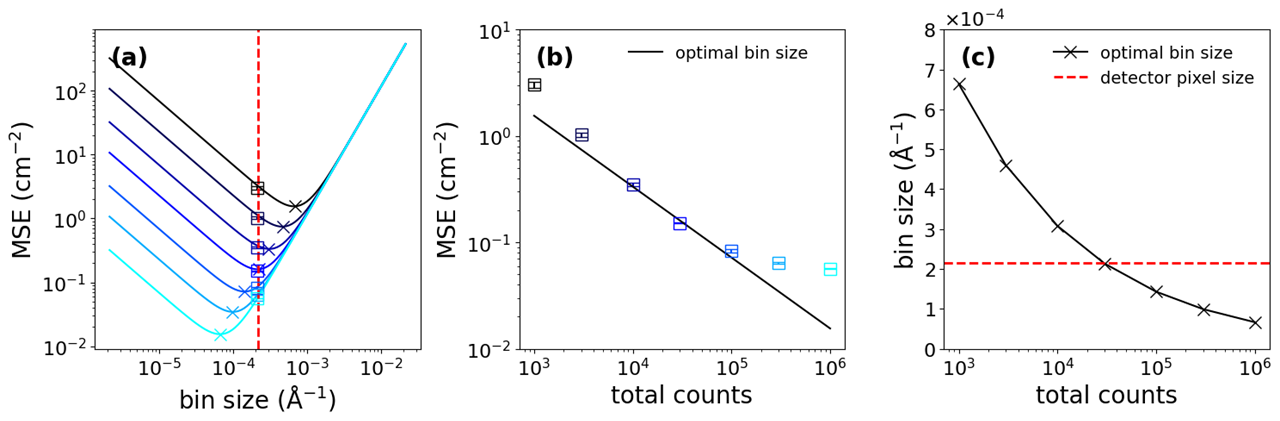}
     \caption{Evolution of the Freedman--Diaconis (FD) optimal binning with total detector counts. (a) Mean-squared error (MSE) versus bin width $h$ for increasing total counts, shown from black to blue. The red dashed line marks the detector-pixel width, and black crosses indicate FD-optimal bin sizes. The minima shift from larger to smaller $h$ as counts increase, marking a transition from the noise-limited to the resolution-limited regime. (b) Minimum MSE values as a function of total counts. The solid line shows the FD-predicted scaling $\langle (\Delta I)^2 \rangle_{\mathrm{opt}} \propto N_{\mathrm{total}}^{-1}$, while open squares correspond to pixel-sized bins. (c) FD-optimal bin width $h_{\mathrm{opt}}$ versus total counts. The black crosses follow the expected scaling $h_{\mathrm{opt}} \propto N_{\mathrm{total}}^{-1/3}$, with the red dashed line denoting the pixel width. The crossover point marks where detector resolution begins to limit statistical improvement.
}
    \label{fig:2}
\end{figure}

Building on the demonstration in Fig.~\ref{fig:1} for a fixed total count, Fig.~\ref{fig:2} extends the analysis to explore how the optimal bin width evolves as the total detector counts increase. Whereas Fig.~1 illustrates the interplay between counting noise and binning distortion for a single experimental condition, Fig.~\ref{fig:2} investigates this competition across multiple statistical regimes, ranging from low-count, noise-dominated conditions to high-count, resolution-limited measurements. This analysis directly connects the Freedman--Diaconis optimization to realistic SANS acquisition scenarios where the total number of detected neutrons varies with sample contrast, detector efficiency, and beamtime.

Fig.~\ref{fig:2} extends the analysis of Fig.~\ref{fig:1} by examining how the optimal binning behavior evolves with increasing total detector counts. While Fig.~\ref{fig:1} illustrated the balance between statistical noise and binning distortion for a single dataset, Fig.~\ref{fig:2} explores this balance systematically across different counting regimes, thereby establishing how statistical quality dictates the optimal resolution achievable in SANS measurements.

Panel~(a) shows the mean-squared error (MSE) between the reconstructed and reference intensity profiles as a function of bin width $h$ for datasets with progressively increasing total neutron counts. The curves, shaded from black through gray to blue, correspond to increasing count levels, while the red dashed vertical line marks the physical width of a detector pixel. Each curve exhibits a characteristic U-shape, reflecting the trade-off between counting noise, which dominates at small $h$, and binning distortion, which dominates at large $h$. The minima of these curves (black crosses) represent the Freedman--Diaconis (FD) optimal bin widths. At low count levels (black and dark gray curves), the MSE minimum occurs at a bin width larger than the detector pixel size, indicating that coarse binning effectively reduces statistical noise in the count-limited regime. As the total counts increase (gray to blue curves), the MSE minima shift systematically toward smaller $h$, demonstrating that improved counting statistics allow finer binning. Once the optimal $h$ becomes comparable to or smaller than the physical pixel size, the analysis enters the resolution-limited regime, where detector geometry rather than statistical uncertainty defines the achievable precision. The intersection between the MSE minima and the red dashed line thus marks the crossover between the noise- and resolution-limited regimes.

Panel~(b) provides a complementary view of these results by plotting the minimum MSE values (from panel~a) as a function of total detector counts. The solid black line represents the MSE at the FD-optimal bin widths, while the open squares correspond to the MSE obtained when the bin width is fixed at the detector pixel size. The FD-optimal MSE decreases approximately as $\langle (\Delta I)^2 \rangle_{\mathrm{opt}} \propto N_{\mathrm{total}}^{-1}$, consistent with the inverse scaling of counting noise in Eq.~\eqref{eq:DeltaI_count}. At low counts, the MSE for the fixed pixel size closely follows the FD-optimal trend, showing that pixel-level binning is nearly optimal when statistical noise dominates. However, as the total counts increase, the fixed-pixel MSE deviates upward, while the FD-optimal MSE continues to decrease, indicating that finer binning becomes advantageous once sufficient signal statistics are available.

Panel~(c) directly relates the FD-optimal bin width to total detector counts. The optimal $h$ (black crosses) decreases monotonically with increasing counts, following the expected FD scaling $h_{\mathrm{opt}} \propto N_{\mathrm{total}}^{-1/3}$. The red dashed horizontal line marks the detector pixel size, highlighting the same crossover observed in panels~(a) and~(b). For $N_{\mathrm{total}} \lesssim 10^{4}$, the optimal bin width is substantially larger than the pixel size, reflecting the dominance of statistical fluctuations. As $N_{\mathrm{total}}$ increases, the optimal width approaches the pixel size around $N_{\mathrm{total}} \approx 10^{4}$–$10^{5}$, and for higher counts, it becomes smaller than the pixel size, signifying the onset of detector-limited performance.

Together, the three panels in Fig.~\ref{fig:2} delineate two distinct operating regimes. In the noise-limited regime, coarse FD binning improves statistical reliability by suppressing counting fluctuations, whereas in the resolution-limited regime, the physical pixel size constrains the achievable precision. The smooth evolution of both MSE and optimal bin width with total counts demonstrates that the Freedman--Diaconis framework provides a unified, statistically grounded criterion that consistently links measurement statistics, instrumental resolution, and data representation in SANS experiments.

\section{Conclusions}
\label{sec:4}

We have developed a statistically rigorous binning framework for Small-Angle Neutron Scattering (SANS) based on the Freedman–Diaconis (FD) rule, providing a quantitative link between measurement statistics, instrumental resolution, and data representation. By deriving explicit scaling relations for counting noise and binning distortion, we established an optimal bin width that balances statistical precision against structural fidelity. Application to model SANS data confirmed that the FD criterion identifies a well-defined minimum in total mean-squared deviation, accurately reproducing the intrinsic curvature of $I(Q)$ while suppressing stochastic fluctuations. The predicted scaling, $h_{\mathrm{opt}} \propto N_{\mathrm{total}}^{-1/3}$, was verified across multiple statistical regimes, delineating the crossover between noise- and resolution-limited behavior. It is worth noting that the scaling law is independent of the choice of variable, making it applicable to both uniformly spaced detector angles and logarithmically spaced bins. This framework offers a unified, physics-informed foundation for adaptive histogramming in neutron scattering, enabling consistent and reproducible data reduction across diverse instruments and sample conditions. Future work will extend this approach to multi-dimensional detector data and incorporate spatial correlations in scattering intensity.

\begin{acknowledgments}
This work was conducted at the Spallation Neutron Source, a U.S. Department of Energy Office of Science User Facility operated by Oak Ridge National Laboratory. JEA was supported by the Next Generation STEM Internship (NGSI) Program at Oak Ridge National Laboratory.
\end{acknowledgments}

%

\end{document}